\documentclass[oribibl,fleqn]{llncs}
\pdfoutput=1
\usepackage[utf8]{inputenc}
\usepackage{hyperref}
\usepackage{ifthen}
\usepackage{graphicx}
\usepackage{isabelle,isabellesym}\isabellestyle{it}

\isadroptag{theory}

\renewcommand{\isastyletxt}{\isastyletext}

\begin{document}

\title{Isabelle/jEdit --- a Prover IDE \\ within the PIDE framework}
\author{Makarius Wenzel}
\institute{Univ. Paris-Sud, Laboratoire LRI, UMR8623, Orsay, F-91405, France \\
CNRS, Orsay, F-91405, France \\
\url{http://www.lri.fr/~wenzel/}}
\maketitle

\begin{isabellebody}%
\def\isabellecontext{Paper}%
\isadelimtheory
\endisadelimtheory
\isatagtheory
\isacommand{theory}\isamarkupfalse%
\ Paper\isanewline
\isakeyword{imports}\ Main\isanewline
\isakeyword{begin}%
\endisatagtheory
{\isafoldtheory}%
\isadelimtheory
\endisadelimtheory
\isamarkupsection{Overview%
}
\isamarkuptrue%
\begin{isamarkuptext}%
PIDE is a general framework for document-oriented prover
  interaction and integration, based on a bilingual architecture that
  combines ML and Scala \cite{Scala:2004}.  The overall aim is to
  connect LCF-style provers like Isabelle \cite[\S6]{Wiedijk:2006} (or
  Coq \cite[\S4]{Wiedijk:2006} or HOL \cite[\S1]{Wiedijk:2006}) with
  sophisticated front-end technology on the JVM platform, overcoming
  command-line interaction at last.

  The present system description specifically covers Isabelle/jEdit as
  part of the official release of Isabelle2011-1 (October 2011).  It
  is a concrete Prover IDE implementation based on Isabelle/PIDE
  library modules (implemented in Scala) on the one hand, and the
  well-known text editor framework of jEdit (implemented in Java) on
  the other hand.

  The interaction model of our Prover IDE follows the idea of
  continuous proof checking: the theory source text is annotated by
  semantic information by the prover as it becomes available
  incrementally.  This works via an asynchronous protocol that neither
  blocks the editor nor stops the prover from exploiting parallelism
  on multi-core hardware.  The jEdit GUI provides standard metaphors
  for augmented text editing (highlighting, squiggles, tooltips,
  hyperlinks etc.) that we have instrumented to render the formal
  content from the prover context.  Further refinement of the jEdit
  display engine via suitable plugins and fonts approximates
  mathematical rendering in the text buffer, including symbols from
  the {\TeX} repertoire, and sub-/superscripts.

  Isabelle/jEdit is presented here both as a usable interface for
  current Isabelle, and as a reference application to inspire further
  projects based on PIDE.%
\end{isamarkuptext}%
\isamarkuptrue%
\isamarkupsection{Using the System%
}
\isamarkuptrue%
\begin{isamarkuptext}%
The described system is part of the official release
  Isabelle2011-1 (October 2011).  The download archives from
  \url{http://isabelle.in.tum.de/website-Isabelle2011-1/download.html}
  cover the three main platform families: Linux, Mac OS X, and Windows
  (with Cygwin).  Isabelle/jEdit has already a history of about 4
  years; a preliminary version is discussed in \cite{Wenzel:2010}.
  October 2011 marks the point of the first stable release of the
  Prover IDE; some remaining limitations are described in its README
  panel.  The website \url{http://isabelle.in.tum.de} points
  dynamically to the latest official release, and further improvements
  of Isabelle/jEdit can be anticipated with coming Isabelle
  distributions.

  The Isabelle distribution bundles sources and multi-platform
  binaries, including Isabelle/jEdit.  Conceptually, the Prover IDE is
  a \emph{rich-client platform} with significant hard-disk foot-print,
  but it runs seamlessly for most users.  The shell command
  \texttt{Isabelle2011-1/bin/isabelle jedit} opens a text editor
  session of jEdit which we have augmented by some plugins to
  communicate with the prover in the background.  Source files with
  \texttt{.thy} extension are treated specifically: Isabelle/jEdit
  adds them to the formal document-model of Isabelle/PIDE, that
  maintains semantic information provided by the prover in the
  background, while the user is editing the text in the foreground.

  The subsequent screenshot shows the editor view after opening a
  certain \texttt{Example.thy} file.  The Isabelle distribution
  contains many other examples, e.g.\
  \texttt{Isabelle2011-1/src/HOL/Unix/Unix.thy} where the editor will
  also propose to load further imported theory files.

  \begin{center}
  \includegraphics[width=\textwidth]{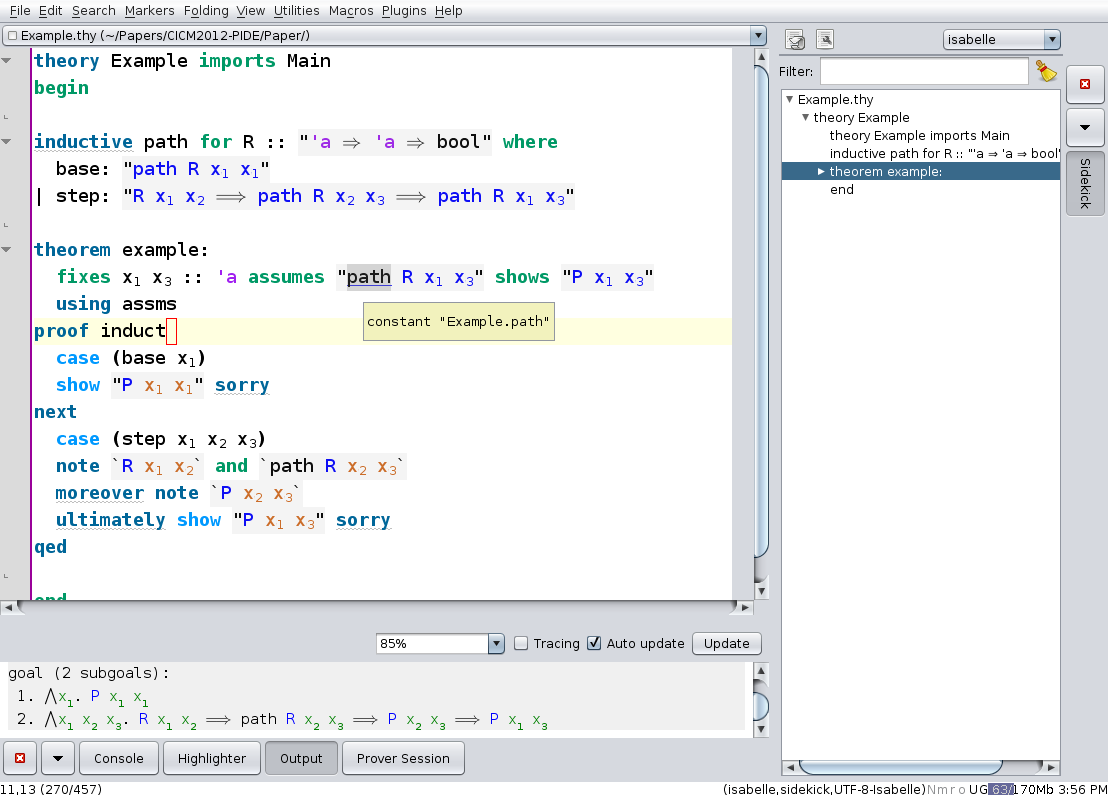}
  \end{center}

  \noindent The main text area is surrounded by \emph{dockable
  windows} that are associated with jEdit plugins.  For example, we
  provide \emph{Output} for prover messages and traditional goal
  states, which is internally based on existing HTML4/CSS2 rendering
  on the JVM.  The tree view is provided by \emph{Sidekick}, which is
  an existing jEdit plugin that has been instrumented to understand
  some Isabelle theory structure.

  \medskip The general aim of Isabelle/jEdit is to expose the specific
  virtues of both Isabelle and jEdit to the user, without accidental
  technical limitations imposed on either system.  This is in contrast
  to the classic Proof~General~/ Emacs \cite{Aspinall:TACAS:2000},
  where the \emph{locked region} is essentially an intrusion of the
  prover command prompt into the editor; it restricts the user to a
  single focus inherited from TTY mode.

  Having replaced the prover Read-Eval-Print-Loop by native document
  editing in Isabelle/PIDE, we can connect the editor more directly.
  The sophisticated features that qualify jEdit as ``Programmer's Text
  Editor''\footnote{\url{http://www.jedit.org}} are retained, and
  augmented by the semantic information from the prover.  The
  underlying JVM platform is sufficiently flexible to support our
  requirements for this formal document-model, but instead of Java we
  are always using Scala \cite{Scala:2004} for our own
  implementations.  Higher-order functional-object-oriented
  programming on multi-threaded JVM is far removed from untyped
  single-threaded Emacs-Lisp.

  \medskip Physical rendering of document content draws from the
  standard repertoire of known IDEs for programming languages, with
  highlighting, squiggles, tooltips, hyperlinks etc.  In the above
  screenshot, only the bold keywords of the Isar language use
  traditional syntax-highlighting in jEdit with static tables; all
  other coloring is based on dynamic information from the logical
  context of the prover.

  Such annotated text regions can be explored further by using the
  \texttt{CONTROL} modifier key (or \texttt{COMMAND} on Mac OS X),
  together with mouse hovering or clicking.  It reveals tooltips and
  hyperlinks, e.g.\ see \texttt{constant "Example.path"} above, and
  thus explains how a certain piece of source text has been
  interpreted.

  \medskip The combination of Isabelle/jEdit and the underlying
  semantic document-model should help users that are accustomed to
  Netbeans or Eclipse to approach formal logic and formalized
  mathematics.  Thus we hope to see new generations of users
  continuing the tradition of the ``LCF approach'' from the 1970-ies.%
\end{isamarkuptext}%
\isamarkuptrue%
\isamarkupsection{Implemented concepts%
}
\isamarkuptrue%
\begin{isamarkuptext}%
Conceptually, the implementation consists of two main parts:
  (1) Isabelle/PIDE infrastructure in ML and Scala that is considered
  an integral part of the prover distribution, and (2) Isabelle/jEdit
  plugins and supporting code to assemble the main application.  PIDE
  provides the main concepts for document-oriented interaction, and is
  most challenging to implement.  Some aspects of previous versions
  are described in \cite{Wenzel:2010,Wenzel:2011:CICM}, but the main
  issues are still unpublished.  Compared to that the jEdit
  application is relatively small and simple: \isa{{\isaliteral{5C3C617070726F783E}{\isasymapprox}}} 100\,Kb of
  Scala code.

  The implemented concepts of Isabelle/jEdit in Isabelle2011-1 that
  are visible to end-users are as follows:
  \begin{description}

  \item [Continuous checking] of source text while editing; no locking,
  no need to save intermediate files.

  \item [Dependency management] between text buffers: each theory file
  corresponds to a node in the development graph of the current
  Isabelle session.  Imports are resolved by reloading required files;
  edits on some node are propagated through the dependency graph as
  expected.

  \emph{Limitation:} non-theory add-on files still need to be managed
  manually, to ensure that the prover loads the proper version.

  \item [Status overview] of single text buffers and the overall
  prover session, with incremental update while the prover processes
  theories and proofs (usually in parallel on multi-core hardware).

  \item [Annotated input] of source text, which is semantically
  decorated and physically highlighted via standard GUI metaphors.

  \item [Annotated output] of prover messages, which is produced by
  traditional pretty-printing of the term language that is augmented
  by semantic markup.  Rendering is delegated to HTML4/CSS on the JVM.

  \emph{Limitation:} no hyperlinks within the browser window yet.

  \item [Integration of Isabelle/ML] into the Prover IDE: ML source
  inside Isar is fully annotated by the compiler, with inferred types
  and identifier scopes.

  \item [Integration of Isabelle/Scala] into the jEdit \emph{Console}
  plugin, which provides command line to access the running JVM via
  the Scala toplevel.

  \emph{Limitation:} only minimal IDE integration via terminal window.

  \item [Mathematical rendering] of the source text based on Unicode
  characters, custom-made \emph{IsabelleText} font with common glyphs
  from the {\TeX} repertoire, and sub-/superscripts via extended jEdit
  text styles

  \emph{Limitation:} only 1-dimensional layout following traditional
  text editing, no support for 2-dimensional boxes (fractions, roots,
  matrices).

  \item [Completion mechanism] for mathematical symbols and keywords
  of the formal Isar language.

  \emph{Limitation:} based on static tables, no connection to semantic
  context yet.

  \end{description}

  \noindent Regular jEdit functionality and generic plugin can be used
  as well.  The physical representation of formal sources coincides
  with JVM and jEdit conventions.  So copy-and-paste or hyper-search
  of mathematical symbols does not cause any surprises to jEdit users.%
\end{isamarkuptext}%
\isamarkuptrue%
\isadelimtheory
\endisadelimtheory
\isatagtheory
\isacommand{end}\isamarkupfalse%
\endisatagtheory
{\isafoldtheory}%
\isadelimtheory
\endisadelimtheory
\isanewline
\end{isabellebody}%
%%% Local Variables:
%%% mode: latex
%%% TeX-master: "root"
%%% End:

\bibliographystyle{abbrv}
\bibliography{root}

\end{document}